\documentclass[aps,prd,final,twocolumn,floats,floatfix,nofootinbib,10pt]{revtex4-1}

\usepackage{graphicx}
\usepackage{amsmath}
\usepackage{amsfonts}
\usepackage{amssymb}
\usepackage{amstext}
\usepackage[english]{babel}
\usepackage{enumitem}
\usepackage{helvet}
\usepackage{microtype}
\usepackage[pdftex]{hyperref}
\usepackage{multirow}
\usepackage{subcaption}
\usepackage{float}

\begin{document}
\title{New Circuit for Quantum Adder by Constant}

\author{Dmytro Fedoriaka}

\date{January 13, 2025}

\begin{abstract}
We propose a new circuit for in-place addition of a classical $n$-bit constant to a quantum $n$-qubit integer modulo $2^n$. Our circuit uses $n-3$ ancilla qubits and has a T-count of $4n-5$. We also propose controlled version of this circuit that uses $n-2$ ancillas and has a T-count of $11n-15$. We implement these circuits in Q\#.
\end{abstract}

\maketitle

\section{Introduction}

Addition by a constant is a building block for various quantum arithmetic algorithms. In particular, it is used to implement a modular exponentiation circuit for the Shor algorithm \cite{liu2021cnot}.

\begin{figure}[H]
  \centering
  \includegraphics[width=8cm]{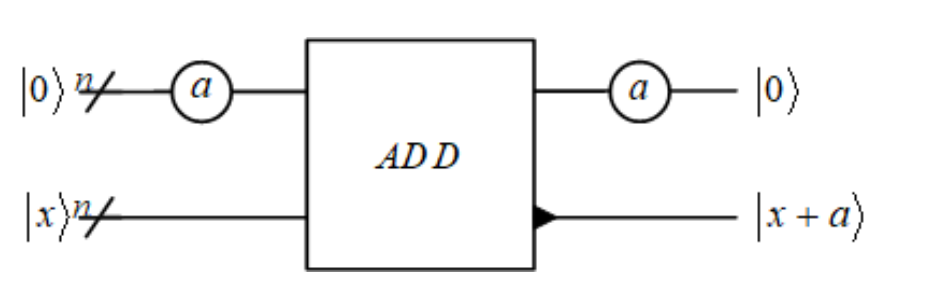}
  \caption{Quantum adder by constant built from in-place quantum adder (figure from \cite{liu2021cnot}) }
  \label{fig:reduction_adder}
\end{figure}

The traditional approach for building an adder-by-constant is to reduce this problem to in-place addition of two quantum numbers: allocate $n$ ancillas, populate them with value of the constant (using X gates), use some in-place quantum adder (e.g. one of \cite{cuccaro2004new,draper2004logarithmic,takahashi2009quantum,gidney2018halving}) to perform the addition of 2 quantum numbers, then uncompute the constant (as shown in fig. \ref{fig:reduction_adder}). This is what is proposed in \cite{liu2021cnot}, and this is how Q\#'s \texttt{Std.Arithmetic.IncByL} is implemented \cite{qsharpStd}.

The other approach is QFT-based \cite{draper2000addition,pavlidis2014fast}, which does not require any ancilla, but relies on arbitrarily precise rotations.

In this paper, we show that it is possible to implement addition-by-constant more efficiently (reducing the number of ancillas and T-count). We do this by generalizing the incrementer circuit from \cite{li2014class} to handle addition by any constant, and then using the $AND$ gate proposed in \cite{gidney2018halving} to reduce the T-count.

\section{Problem statement and notation}

\textbf{Problem statement.} Let $a = \sum_{i=0}^{n-1} (a_i \cdot 2^i)$ --- a classical constant. Let $b = \sum_{i=0}^{n-1} (b_i \cdot 2^i)$ --- a number encoded in a quantum register $| b \rangle$. Let $s=(a+b) \mod 2^n=\sum_{i=0}^{n-1} (s_i \cdot 2^i)$. Our task is to build a quantum circuit that acts on a register of $n$ qubits and applies an unitary matrix $U_{+a}$, such that $U_{+a} | b \rangle = | s \rangle.$

\textbf{Classically controlled X gate.} We define ``classically controlled X gate", acting on qubit $q$, controlled on classical bit $a$, as $X(q)$ if $a=1$ and as $I(q)$ (no-op) if $a=0$. We denote this $X_a(q)$. In a circuit diagram, we denote it by ``$a$" in a circle, as shown on fig. \ref{fig:temp_logical_and}. 

\begin{figure}
  \centering
  \includegraphics[width=6cm]{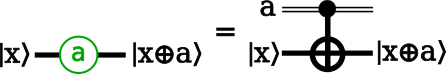}
  \caption{Classically controlled X gate}
  \label{fig:temp_logical_and}
\end{figure}

\textbf{AND gate.} We use ``temporary logical-$AND$" gate, or simply ``$AND$ gate", as introduced in \cite{gidney2018halving} by Craig Gidney. This gate always comes in a pair with its adjoint $AND ^ \dag$. This gate is equivalent to the Toffoli gate, except the target qubit must be in state $| 0 \rangle$ before applying the $AND$ gate. As shown in \cite{gidney2018halving}, T-cost of a pair of $AND$ and $AND^\dag$ gates is 4, while T-cost of a single Toffoli gate is 7. Notation for $AND$ and $AND^\dag$ gates is on fig. \ref{fig:temp_logical_and}.

\begin{figure}
  \centering
  \includegraphics[width=3cm]{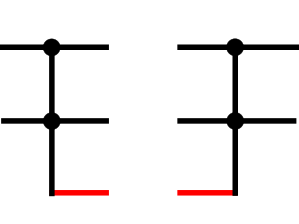}
  \caption{Temporary logical-$AND$ gate and its adjoint}
  \label{fig:temp_logical_and}
\end{figure}

\textbf{MAJ gate.} We use the following definition of the ``majority" function: $MAJ(x,y,a)=\lfloor (x+y+a)/2 \rfloor$. This can be rewritten as:

$$
MAJ(x, y, a) = 
\begin{cases}
 x \wedge y, \text{if}~ a=0 \\
 x \vee y, \text{if}~ a=1
\end{cases}
=
((x \oplus a) \wedge (x \oplus a)) \oplus a.
$$

Here we used the de Morgan's law: $x \vee y = ((x \oplus 1) \wedge (x \oplus 1)) \oplus 1$.

\begin{figure}
  \centering
  \includegraphics[width=5cm]{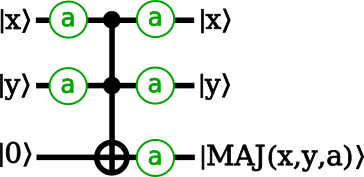}
  \caption{Implementing $MAJ_a$ gate}
  \label{fig:MAJ_gate}
\end{figure}

Now, it's easy to see how to implement gate $MAJ_{a}$, such that $MAJ_{a} | x \rangle | y \rangle | 0 \rangle =| x \rangle | y \rangle | MAJ(x,y,a) \rangle$. This can be done using one Toffoli gate and 5 $X_a$ gates, as shown on fig. \ref{fig:MAJ_gate}.

Throughout this paper, we denote input/output qubits with black lines and ancillary qubits with red lines. 

When estimating circuit complexity, we will assume $n \ge 4$.

\section{Circuit construction}

\subsection{The basic idea}

First, note that if $a$ is even, then $a=2^z \cdot a'$ where $a'$ is odd, so the task is reduced to adding number $a'$ to $n-z$ most significant bits of $b$. If $a=0$, the adder is an empty circuit. If $a \notin [0, 2^n-1]$, the task is reduced to adding $a \mod 2^n$. So, moving forward, we can assume that $a \in [1, 2^n-1]$ and $a$ is odd (that is, $a_0=1$). 

We use the standard addition with carry: $s_i = a_i \oplus b_i \oplus c_i$, where $c_i$ are the carry bits. The carry bits are computed as follows: $c_0=0$ (no input carry); $c_1=MAJ(a_0, b_0, c_0) = MAJ(1, b_0, 0)=b_0$. For $i=2 \dots n-1$: $c_{i} = MAJ(a_{i-1}, b_{i-1}, c_{i-1})$.

Let's start with computing $s_i = a_i \oplus b_i \oplus c_i$ for $i=n-1$. We use $n-1$ ancillas to compute $c_1, c_2, \dots c_{n-1}$ using $MAJ_{a_i}$ gates, then add CNOT($c_{n-1}$, $b_i$), and uncompute all the carries. Finally, we add $X_{a_{n-1}}(b_{n-1})$.

We repeat this for $i=n-1, n-2, \dots 1$. Finally, we add $X_{a_0}(b_0) = X(b_0)$ to correctly compute $s_0= b_0 \oplus a_0 = b_0 \oplus 1$. The resulting circuit (for $n=5$) is shown on fig. \ref{fig:adder1}.

\subsection{Optimizing to $O(n)$ gates}

\begin{figure*}
  \centering
  \includegraphics[width=18cm]{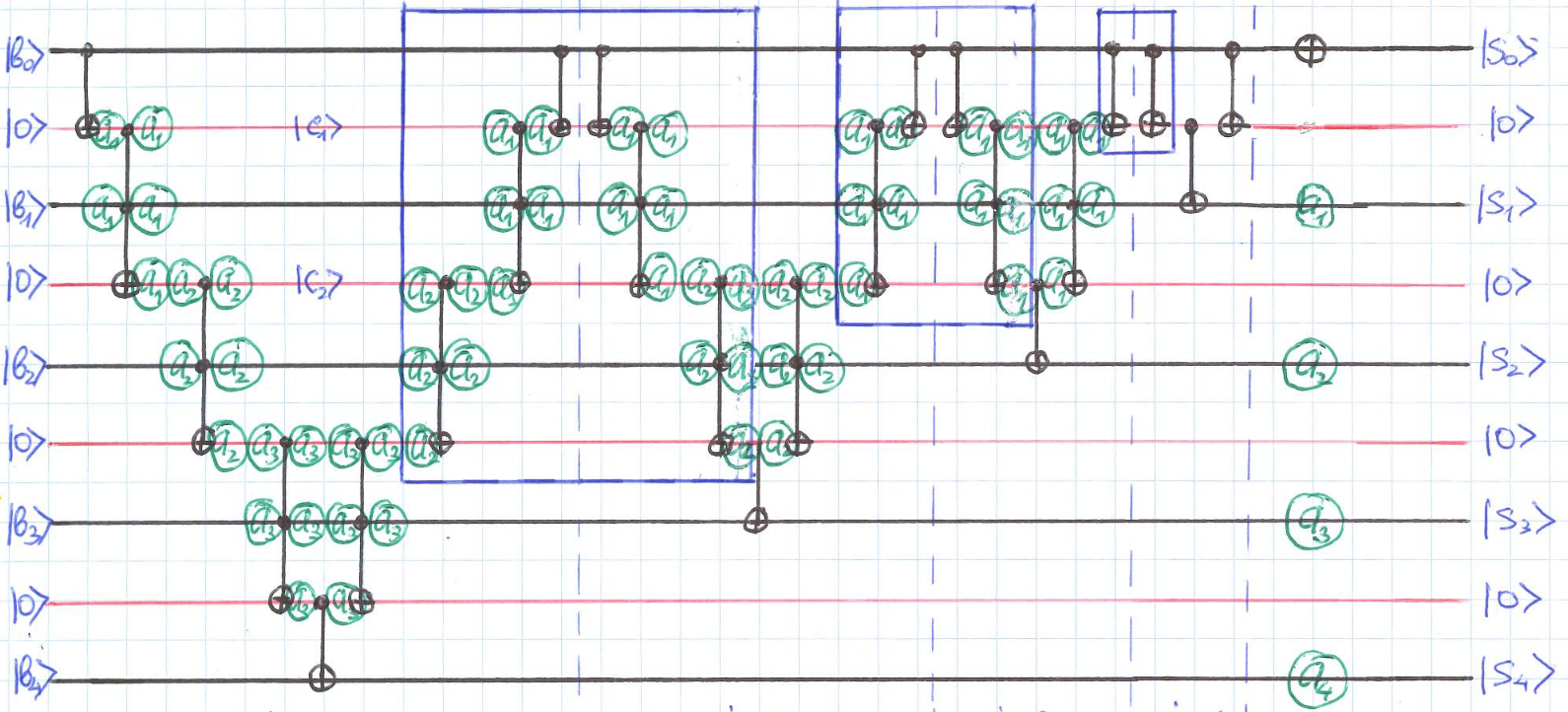}
  \caption{5-qubit Quantum Adder by Constant (unoptimized)}
  \label{fig:adder1}
\end{figure*}

\begin{figure*}
  \centering
  \includegraphics[width=18cm]{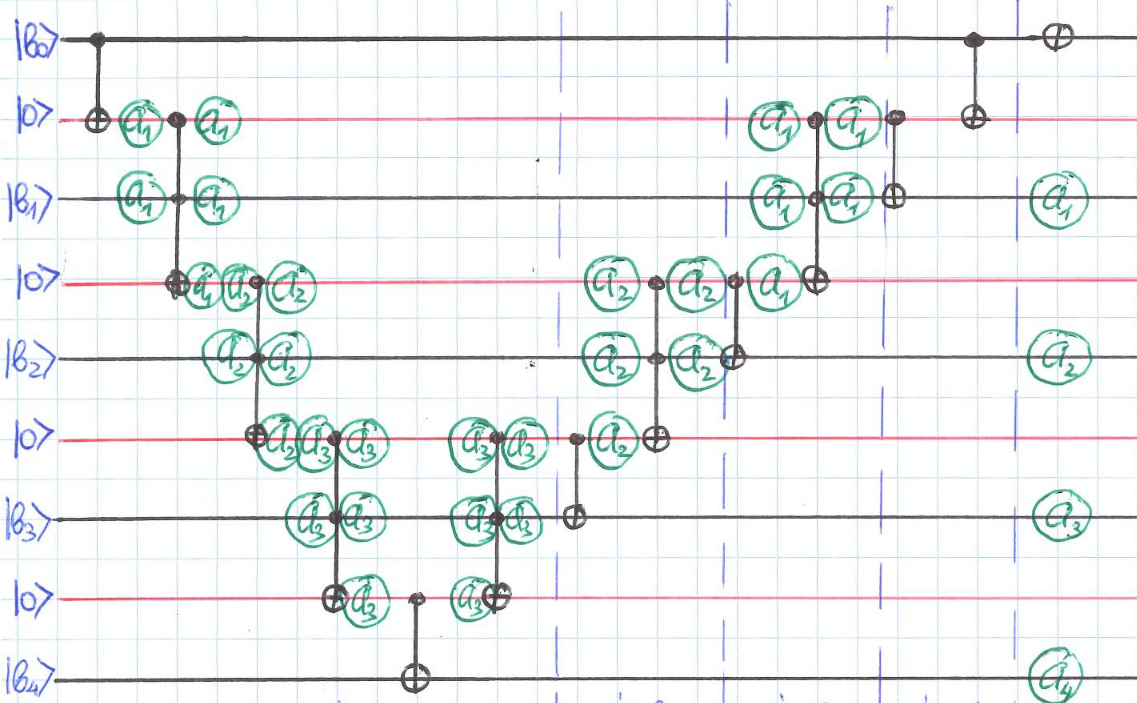}
  \caption{5-qubit Quantum Adder by Constant (partially optimized)}
  \label{fig:adder2}
\end{figure*}

We notice that some adjacent blocks of gates are adjoint of each other and can be canceled. These blocks are shown in blue rectangles on fig. \ref{fig:adder1}. The circuit after cancellation, as shown in fig. \ref{fig:adder2}, has $O(n)$ gates.

This is very similar to the construction of a quantum incrementer (adder by 1) in \cite{li2014class}. Our circuit without classically controlled X gates is identical to their construction of the incrementer.

\subsection{Further optimizations}

Now, we do more optimizations:
\begin{enumerate}[noitemsep,topsep=0pt]
    \item Eliminate adjacent classically controlled X gates.
    \item Merge adjacent gates $X_{x}(q)$ and  $X_{y}(q)$ into $X_{x \oplus y}(q)$.
    \item Eliminate the first ancilla, because $c_1=b_0$.
    \item Cancel some $X_{a_i}(b_i)$ gates in the last column.
    \item Push two $X_{a_{n-2}}(c_{n-1})$ gates to the qubit $b_{n-1}$ and merge it with $X_{a_{n-1}}(b_{n-1})$ gate, resulting in the $X_{a_{n-2} \oplus a_{n-1}}(b_{n-1})$ gate.
    \item Eliminate the last ancilla and one Toffoli gate, by acting with the last Toffoli directly on the $b_{n-1}$.
    \item Replace the Toffoli gates with $AND$ and $AND^{\dag}$ gates.
\end{enumerate}

\begin{figure*}
  \centering
  \includegraphics[width=14cm]{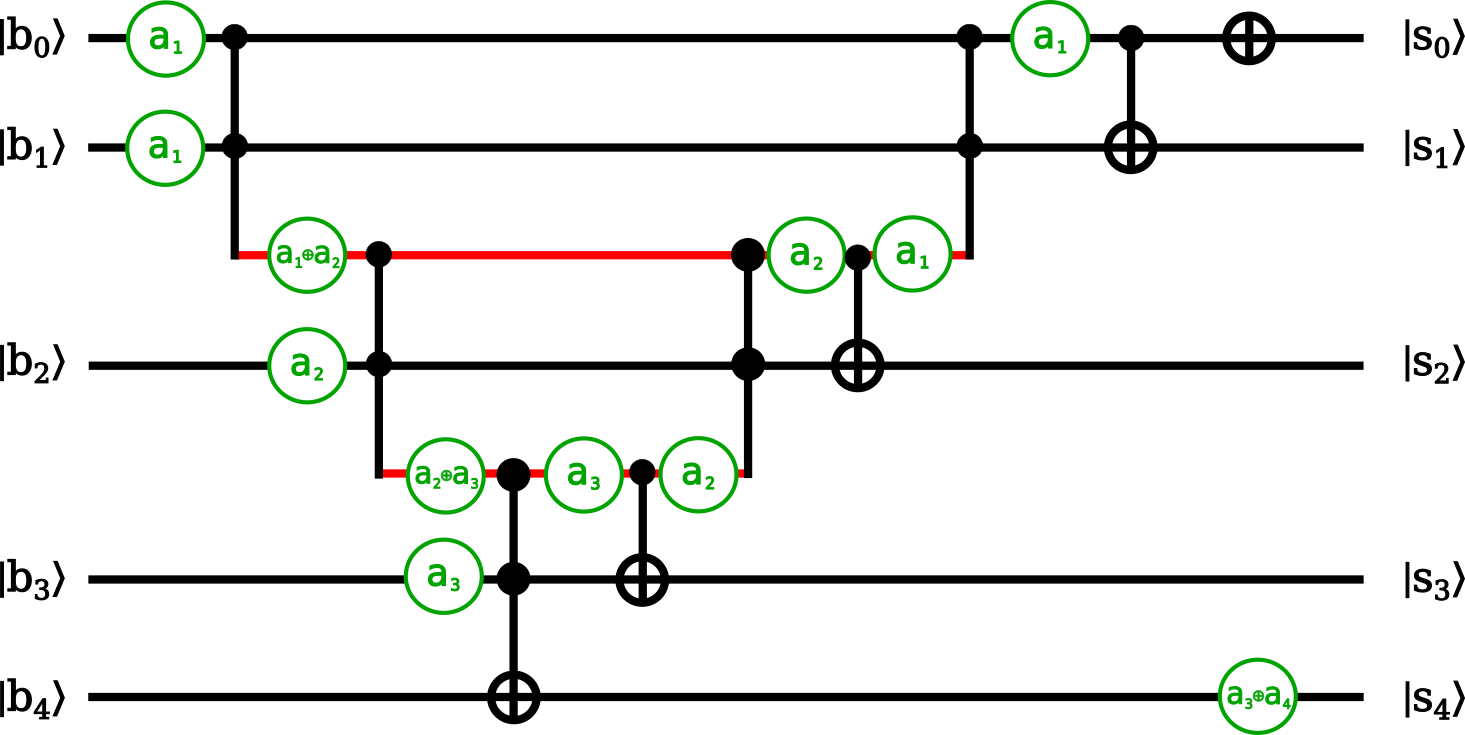}
  \caption{5-qubit Quantum Adder by Constant (optimized)}
  \label{fig:adder3}
\end{figure*}

The optimized circuit (shown on fig. \ref{fig:adder3} for $n=5$) contains $4n-8$ classically controlled $X$ gates, $n-3$ pairs of $AND$ and $AND ^\dag$ gates, 1 Toffoli gate, $n-2$ CNOT gates and one $X$ gate. It uses $n-3$ ancillas and has T-count of $4 \cdot (n-3)+7 = 4n-5$.

\subsection{Controlled version}

\begin{figure*}
  \centering
  \includegraphics[width=14cm]{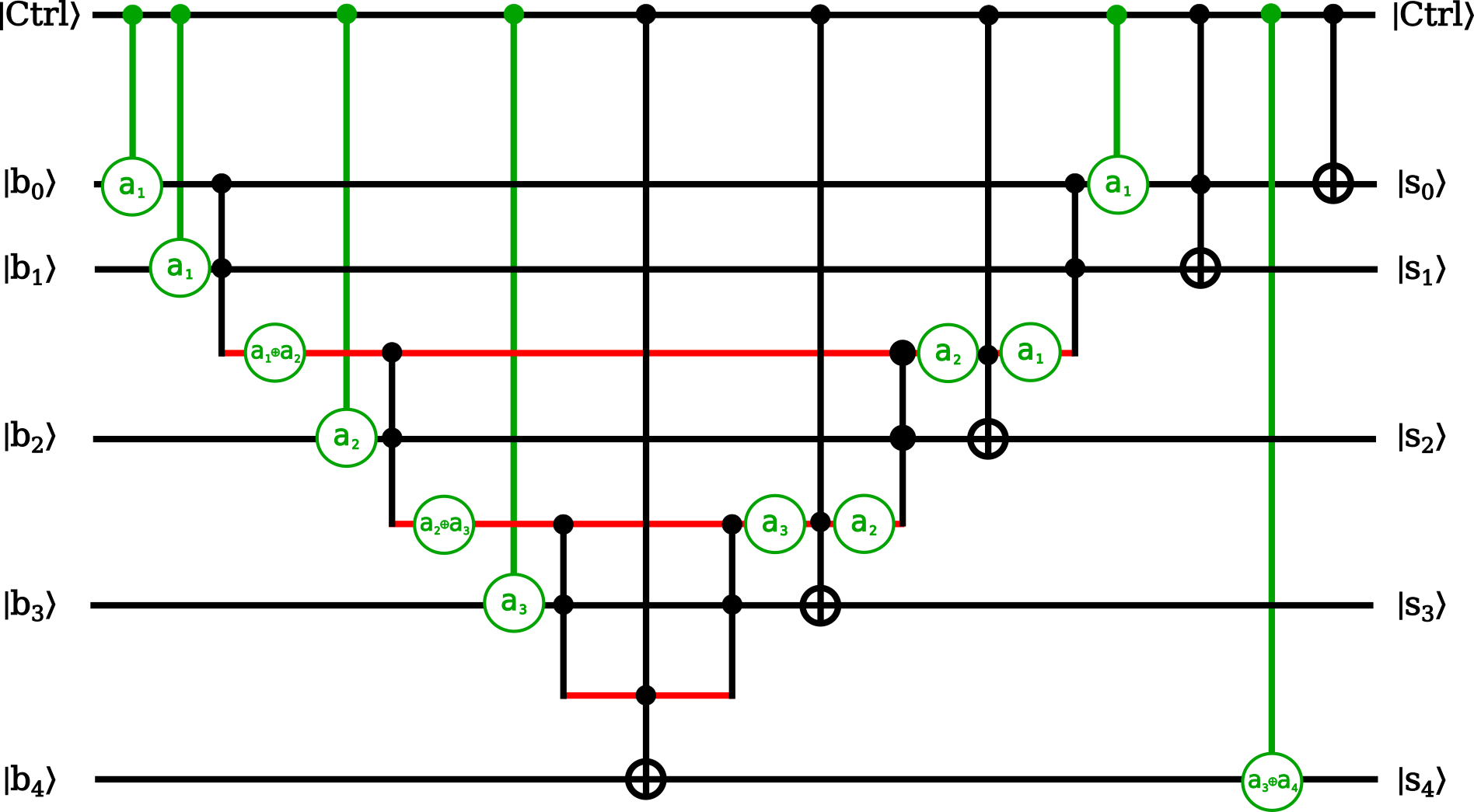}
  \caption{5-qubit Controlled Quantum Adder by Constant}
  \label{fig:adder4}
\end{figure*}

Proposed adder can be extended to controlled adder-by-constant. For that, we need to make all the gates that target input/output qubits (but not ancillas) controlled. We can keep all optimizations except 3.3.6 (where we eliminated the last ancilla), because if we do this optimization, we would need a ``CCCNOT" gate in the controlled circuit. Fig. \ref{fig:adder4} shows circuit diagram for the controlled adder-by-constant for $n=5$.

Our controlled adder-by-constant uses $n-2$ ancillas, $n-2$ pairs of $AND$ and $AND ^\dag$ gates, and $n-1$ Toffoli gates. T-count of this circuit is $4\cdot(n-2) + 7 \cdot(n-1)=11n-15$.

\subsection{Implementation}

We implemented the proposed circuit (both uncontrolled and controlled version) in the Q\# language in \cite{constAdderImpl}. Also, there one can find comparison-by constant circuits built using a similar idea.

\section{Complexity analysis}

We compare our circuit with several known in-place quantum adders: Cuccaro Ripple-Carry Adder (RCA) \cite{cuccaro2004new}, Draper Carry-Lookahead Adder (CLA) \cite{draper2004logarithmic}, Takahashi RCA \cite{takahashi2009quantum} and Gidney RCA \cite{gidney2018halving}. For these adders we assume that adders-by-constants were built using procedure shown on fig. \ref{fig:reduction_adder}, which adds $n$ ancillas and doesn't affect T-count. We use two metrics: number of ancilla qubits, and T-count (number of T-gates needed to build the circuit). Our comparison results are in table \ref{tab:costs}.

\begin{table}[H]
\centering
\caption{Cost comparison of different adders.}
\label{tab:costs}
\begin{tabular}{|l|l|l|}
\hline & Ancilla & T-count \\ \hline
Cuccaro RCA \cite{cuccaro2004new} & $n+1$  & $14n-21$  \\ \hline
Draper CLA \cite{draper2004logarithmic} & $2n - 2 \lceil \log_2 n \rceil -1$  & $70n-84 \lceil \log_2 n \rceil -42$   \\ \hline
Takahashi RCA \cite{takahashi2009quantum} & $n$ & $14n-7 $  \\ \hline
Gidney RCA \cite{gidney2018halving} & $2n-1$ & $4n-4 $  \\ \hline
Proposed & $n-3 $ & $4n-5$    \\ \hline

\end{tabular}
\end{table}

\section{Conclusion}

We proposed a circuit that adds a classical number to a quantum register in-place, using only $n-3$ ancillas and having T-count of $4n-5$. To our knowledge, it's more efficient than any circuit that can be constructed using any known quantum-by-quantum adder.

\bibliographystyle{utphys} 
\bibliography{references}

\end{document}